\documentclass[12pt]{article}
\usepackage{latexsym,amsfonts,amssymb}
\makeatletter
\@addtoreset{equation}{subsection}
\makeatother

\begin{document}
\begin{center}
{\Large \bf The Principle of Sufficient Reason\\ and Quantum Determinism}
\\[1.5cm]
 {\bf Vladimir S.~MASHKEVICH}\footnote {E-mail:
  Vladimir.Mashkevich100@qc.cuny.edu}
\\[1.4cm] {\it Physics Department
 \\ Queens College\\ The City University of New York\\
 65-30 Kissena Boulevard\\ Flushing\\ New York
 11367-1519} \\[1.4cm] \vskip 1cm

{\large \bf Abstract}
\end{center}
The Principle of Sufficient Reason implies determinism. An explicit indeterministic quantum jump dynamics is constructed, which may be naturally transformed into a deterministic one. A consistent application of the Principle of Sufficient Reason results in a canonical deterministic dynamics.
\newpage
\section*{Introduction}

The advent of quantum physics has introduced into description of the universe a new element---indeterminism. Indeterminism, in turn, implies nonuniversality of the Principle of Sufficient Reason. But the denial of the Principle may result in arbitrariness in arguments. Therefore, the Principle formulated explicitly by Leibnitz [1] has been regarded to be an indispensable law of formal logic by a number of philosophers [2--4].

In physics, the concept of indeterminism, or indeterminacy relates to quantum jump dynamics, namely, reflects the contingency of the results of state vector reduction. In a series of the results, there appear probabilistic regularities, but an individual result is unexplainable, which is in conflict with the Principle of Sufficient Reason, or falls under the Principle of Insufficient Reason.

Dissatisfaction with such a state of affairs has been expressed by some physicists, among them Einstein [5,6] and Dirac [7].

A number of attempts to reintroduce the Principle of Sufficient Reason and, by the same token, determinism into quantum physics have been undertaken via introducing hidden variables (see [8,9] and references therein).

In [10], we have advanced a deterministic quantum jump dynamics based on jump instants with respect to graduated cosmic time.

In this paper, an explicit indeterministic quantum jump dynamics is constructed, and this dynamics may be naturally transformed into a deterministic one. Finally, it is shown that a consistent application of the Principle of Sufficient Reason results naturally in the deterministic dynamics advanced in [10], so that the latter may be regarded as a canonical one.

\section{The Principle of Sufficient Reason,\\ determinism and indeterminism}

\subsection{The Principle of Sufficient Reason}

There are four laws of formal logic. Three Aristotelian laws of identity, contradiction, and excluded middle are applied for finding truths of reason; Leibnitz's Principle of Sufficient Reason is applied for discovering empirical or contingent truths.

The Principle of Sufficient Reason is expressed in the following manner [1]:

``Nothing takes place without a sufficient reason. Which means that nothing takes place without it be possible to he who would sufficiently know the things, to give a reason that would suffice for determining why it is so instead of otherwise...

Although the said reason cannot, most of the time, be known to us...''

It is obvious that the Principle has a direct bearing on quantum jump dynamics.

\subsection{Determinism and indeterminism}

Determinism is the doctrine, or theory that all that happens, every event is completely determined by a necessary previously existing cause.

In the realm of classical physics, the concept of determinism has been expressed by Laplace [11]:

``An intelligence capable of knowing, at a given instant, all the forces and the disposition of all the entities present in nature, and sufficiently profound to submit all these data to analysis, would grasp within the same formula the motions of the largest bodies in the universe and those of the lightest atoms; to her, nothing would be uncertain and the future as well as the past would be present to her eyes.''

Indeterminism is the negation of determinism.

\subsection{Interrelation between the Principle of Sufficient Reason,\\ determinism and indeterminism}

The Principle of Sufficient Reason implies determinism, so that we have the implication
\begin{equation} 
\mathrm{the\;Principle\;of\;Sufficient\;Reason}\Rightarrow\mathrm{determinism}
\end{equation}
from which follows another one:
\begin{equation} 
\mathrm{indeterminism}\Rightarrow \mathrm{the\; Principle\; of\; Insufficient\; Reason}
\end{equation}
But, on the other hand,
\begin{equation} 
\mathrm{determinism}\nRightarrow\mathrm{the\;Principle\;of\;Sufficient\;Reason}
\end{equation}
The Principle is a law of logic, whereas determinism is a doctrine relating to the universe and based on the Principle.

\section{Probability and sequential probability spaces.\\ Random and nonrandom sequences}

\subsection{Probability space}

Properties of quantum jumps are described in terms of probability and randomness, so it is appropriate to recall some basics [12,13].

A probability space is a triplet
\begin{equation} 
(\Omega,\mathcal{F},P)
\end{equation}
where $\Omega$ is the space of elementary outcomes, $\omega\in\Omega$, $\mathcal{F}$ is a $\sigma$-algebra of subsets of $\Omega$, and $P$ is a probability measure on a measurable space ($\Omega,\mathcal{F}$). For an event $C\in\mathcal{F}$, $P(C)$ is the probability of $C$.

\subsection{Sequential probability space}

The sequential probability space based on the space (2.1.1) is the space of sequences of elements of $\Omega$, i.e., elementary outcomes:
\begin{equation} 
(\tilde{\Omega},\tilde{\mathcal{F}},\tilde{P}),\qquad \tilde{\Omega}=\{\tilde{\omega}=
(\omega_{1},\omega_{2},\cdots),\;\omega_{i}\in\Omega,\;i\geq 1\}
\end{equation}
where $\tilde{\mathcal{F}}$ and $\tilde{P}$ are generated by $\mathcal{F}$ and $P$ [12].

\subsection{Random and nonrandom sequences}

Roughly speaking, a sequence $\tilde{\omega}\in\tilde{\Omega}$ is nonrandom if and only if it may be given by an algorithm; otherwise it is random. A very important kind of randomness is 1-randomness [13].

As to the Principle of Sufficient Reason, it has a direct bearing on contingency in physics but is unrelated to randomness in mathematics.

\section{Indeterministic quantum jump sequences}

\subsection{Quantum jump: State vector reduction}

State vector reduction is described as follows:
\begin{equation} 
|\rangle=\sum\limits_{i=1}^{n}c_{i}|i\rangle,\;\langle|\rangle=1,\;
\langle i|i'\rangle=\delta_{ii'},\;\sum\limits_{i=1}^{n}|i\rangle\langle i|=I,\;
n\geq 2
\end{equation}
\begin{equation} 
|\rangle\stackrel{\mathrm{reduction}}{\longrightarrow}|i\rangle\quad \mathrm{with\;
probability}\;p_{i}=|c_{i}|^{2}=|\langle i|\rangle|^{2}
\end{equation}

\subsection{Probability space and sequential space}

The probability space (2.1.1) is discrete:
\begin{equation} 
(\Omega,\mathcal{F},P),\quad \Omega=\{|i\rangle:i=1,2,\cdots,n,\;n\geq 2\}
\end{equation}
$\mathcal{F}$ is the family of all subsets of $\Omega$, $\mathcal{F}=2^{\Omega}$,
\begin{equation} 
P(\{i\})=p_{i}
\end{equation}

The sequential space (2.2.1) is:
\begin{equation} 
(\tilde{\Omega},\tilde{\mathcal{F}},\tilde{P}),\quad\tilde{\Omega}=\{
\tilde{\omega}=(|l_{1}\rangle,|l_{2}\rangle,\cdots),\;|l_{k}\rangle\in\Omega,\;k\geq 1\}
\end{equation}

\subsection{Indeterminism: Random sequences}

In quantum theory, it is assumed that sequences $\tilde{\omega}\in\tilde{\Omega}$ related to quantum jumps are random; this is quantum indeterminism.

Now we quote Einstein [6]:

``That the Lord should play with dice, all right; but that He should gamble according to definite rules, that is beyond me.''

Indeed, for different $|\rangle$ and different $\{|i\rangle\}_{i=1}^{n}$, the rules should be different; this is too much to comprehend.

\section{The indeterministic universe}

\subsection{Distribution function for a discrete probability space}

In the universe, actual quantum jumps do not fall under the pattern analyzed in the previous Section. Now we will develop an adequate treatment. As the starting point, we introduce a distribution function on a discrete probability space.

Consider a discrete probability space:
\begin{equation} 
(\Omega,\mathcal{F},P),\;\Omega=\{\omega_{i}:i=1,2,\cdots,n,\;2\leq n\leq\infty\},\;
\mathcal{F}=2^{\Omega},\;p_{i}=:P(\{\omega_{i}\}),\;\sum\limits_{i=1}^{n}p_{i}=1
\end{equation}

Introduce a random variable
\begin{equation} 
\xi:\Omega\rightarrow \mathbb{R},\;\;\omega_{i}\rightarrowtail \xi(\omega_{i})
\end{equation}
with
\begin{equation} 
\xi(\omega_{1})=p_{1},\;\;\xi(\omega_{2})=p_{1}+p_{2},\cdots,\xi(\omega_{i})=\sum
\limits_{k=1}^{i}p_{k}
\end{equation}
The distribution function is
\begin{equation} 
F_{\xi}(x)=P(\{\omega_{i}:\xi(\omega_{i})\leq x\}),\;\;-\infty<x<\infty
\end{equation}
It holds that
\begin{equation} 
\mathrm{for}\; x<p_{1},\;\;F_{\xi}(x)=0;\quad \mathrm{for}\;x\geq 1,\;\;F_{\xi}(x)=1
\end{equation}
\begin{equation} 
\mathrm{for}\;x\in[\xi(\omega_{i}),\xi(\omega_{i+1})),\;i+1\leq n,\;\;F_{\xi}(x)=\xi(\omega_{i})=\sum
\limits_{k=1}^{i}p_{k}
\end{equation}

\subsection{Unification: Reduction to uniform distribution}

The next step is a unification of probability spaces (4.1.1). Introduce the probability space
\begin{equation} 
(\Omega=[0,1),\mathcal{F}=\mathrm{Borel}\,\,\sigma\mbox{-}\mathrm{algebra}\,B,\,P=\mathrm{the\,
Lebesgue\,measure}\,\lambda)
\end{equation}
A random variable $U$ is said to be uniformly distributed on $[0,1)$ if its distribution function has the form
\begin{equation} 
F_{U}(x)=\left\{
\begin{array}{lcl}
0\;\mathrm{for}\;x\leq 0\\
x\;\mathrm{for}\;0\leq x\leq 1\\
1\;\mathrm{for}\;1\leq x\\
\end{array}
\right.
\end{equation}

There exists a transformation which takes the uniform random variable $U$ into an arbitrary one [14]. In the case of the discrete probability space (4.1.1), the construction is as follows. Put
\begin{equation} 
[0,1)\ni r_{i}=\xi(\omega_{i})=\sum
\limits_{k=1}^{i}p_{k}
\end{equation}
then
\begin{equation} 
p_{i}=r_{i}-r_{i-1},\;\;r_{0}=0
\end{equation}
Thus,
\begin{equation} 
p_{i}=|[r_{i-1},r_{i})|=\lambda([r_{i-1},r_{i}))
\end{equation}
\begin{equation} 
\sum\limits_{i=1}^{n}p_{i}=\sum\limits_{i=1}^{n}\lambda([r_{i-1},r_{i}))=
\lambda\left(\bigcup\limits_{1\leq i\leq n}[r_{i-1},r_{i})\right)=
\lambda([0,1))=1
\end{equation}

\subsection{Quantum jump sequence in the universe}

Now, for a quantum jump sequence in the universe,
\begin{equation} 
(\mathrm{jump_{1}},\mathrm{jump_{2}},\cdots,\mathrm{jump}_{j},\cdots),\;\;1\leq j\leq n,\;\;
n\leq\infty
\end{equation}
the probabilities of outcomes are determined by a sequence of reals
\begin{equation} 
(r_{1},r_{2},\cdots,r_{j},\cdots),\;\;r_{j}\in[0,1)
\end{equation}
\begin{equation} 
\mathrm{outcome\;for\;jump}_{j}\;\mathrm{is}\;|ji\rangle\;\;\mathrm{iff}\;\;
r_{j}\in[\sum\limits_{k=1}^{i-1}p_{jk},\sum\limits_{k=1}^{i}p_{jk})
\end{equation}
where $p_{jk}$ is the probability of outcome $k$, i.e., $|jk\rangle$ for $\mathrm{jump}_{j}$.

Return to the quotation in Subsection 3.3. Now the task for the Lord is considerably simplified. For any jump He has to choose a real $r\in[0,1)$.

The sequence
\begin{equation} 
\widetilde{|j\rangle}=(|ji\rangle)_{i=1}^{n_{j}}
\end{equation}
may be fixed by a natural condition:
\begin{equation} 
p_{j1}\geq p_{j2}\geq\cdots
\end{equation}
If
\begin{equation} 
p_{ji}=p_{ji+1}=\cdots=p_{ji+l}
\end{equation}
then the condition is
\begin{equation} 
p_{ji}>p_{ji+1}>\cdots>p_{ji+l}\;\;\mathrm{for}\;\;t=t_{j}+\Delta t,\;\;\Delta t
\rightarrow 0
\end{equation}
where $t_{j}$ is the instant of $\mathrm{jump}_{j}$.

Again, $\mathrm{jump}_{1}$ is the first jump in the only (or any) cycle of the universe.

\subsection{Quantum jump dynamics. Indeterminism}

So, the treatment of quantum jump dynamics in the universe amounts to this. There is the probability space (4.2.1)
\begin{equation} 
(\Omega=[0,1),\mathcal{F}=B,P=\lambda),\;\;\Omega\ni\omega=r\in[0,1)
\end{equation}
and the related sequential probability space
\begin{equation} 
(\tilde{\Omega},\tilde{\mathcal{F}},\tilde{P}),\;\tilde{\Omega}\ni\tilde{\omega}=
\tilde{r}=(r_{1},r_{2},\cdots,r_{j},\cdots),\;\;1\leq j\leq n,\;\;n\leq\infty
\end{equation}
Quantum jump dynamics is determined by a sequence
\begin{equation} 
\tilde{r}=(r_{j})_{j=1}^{n},\;\;r_{j}\in[0,1),\;\;n\leq\infty
\end{equation}
(If $n<\infty$, the sequence (4.4.3) is called a string.)

Let the sequence $\tilde{r}$ (4.4.3) be random and not given in advance. Then the sequence of quantum jumps
\begin{equation} 
\widetilde{\mathrm{jump}}=(\mathrm{jump}_{j})_{j=1}^{n},\;\;n\leq\infty
\end{equation}
is contingent and, by the same token, indeterministic. (We use terms random and contingent for mathematical and physical objects, respectively.)

\subsection{Invoking the Principle of Sufficient Reason:\\ Determinism via randomness}

Now we invoke the Principle of Sufficient Reason, which implies determinism (1.3.1), i.e., a predictable sequence $\tilde{r}$ (4.4.3) and, by the same token, a predictable quantum jump dynamics. There are two possibilities to obtain determinism on the basis of randomness.

(i) $\tilde{r}$ is preassigned.

(ii) A random $r\in[0,1)$ is given and is expressed in the form of a binary expansion:
\begin{equation} 
r=0.b_{1}b_{2}\cdots,\;\;b_{i}\in\{0,1\}
\end{equation}
Put
\begin{equation} 
r_{1}=0.b_{1}b_{2}\cdots,\;\;r_{2}=0.b_{2}b_{3}\cdots,\cdots,\;\;r_{j}=0.b_{j}b_{j+1}\cdots
\end{equation}
Thus we have obtained a given
\begin{equation} 
\tilde{r}=(0.b_{j}b_{j+1}\cdots)_{j=1}^{n},\;\;n\leq\infty
\end{equation}
and, by the same token, a deterministic quantum jump dynamics.

\subsection{Invoking the Principle of Sufficient Reason once more:\\
Determinism via jump instants \\with respect to graduated cosmic time}

The deterministic construction of the previous Section falls under the Principle of Insufficient Reason: There is no sufficient reason for choosing both $\tilde{r}$ in the case (i) and $r$ (4.5.1) in the case (ii): There are infinitely many possibilities. (This is a corroboration of the statement (1.3.3).) Note that, in the capacity of $r$ in the case (ii), it would be possible to choose Chaitin's random number $\Omega_{\mathcal{U}}$ representing the halting probability of a universal Chaitin computer $\mathcal{U}$ since that number is not ``typically 1-random'' [13]. But the set of different $\Omega_{\mathcal{U}}$ is countably infinite.

We take a recourse to a natural construction advanced in [10]. Time is a dimensional quantity, so that
\begin{equation} 
t=t_{\mathrm{unit}}\tau
\end{equation}
where $t_{\mathrm{unit}}$ is a unit of time and $\tau$ is dimensionless. Let $\tau=0$ relate to the beginning of the only (or any) cycle of the universe. It holds that
\begin{equation} 
\tau=m+r,\;\;m=0,1,2,\cdots,\;\;r\in[0,1)
\end{equation}
Let $\tau_{j}$ be the instant of $\mathrm{jump}_{j}$,
\begin{equation} 
\tau_{j}=m_{j}+r_{j}, \;\;m_{j}=0,1,2,\cdots,\;\;r_{j}\in[0,1)
\end{equation}
Choose that $r_{j}$ in the capacity of $r_{j}$ in $\tilde{r}$ (4.4.3). This choice is natural, so that it falls under the Principle of Sufficient Reason.

A natural choice of $t_{\mathrm{unit}}$ is either
\begin{equation} 
t_{\mathrm{unit}}=t_{\mathrm{Planck}}
\end{equation}
or
\begin{equation} 
t_{\mathrm{unit}}=(\varkappa\Lambda) t_{\mathrm{Planck}}
\end{equation}
where $\varkappa=(t_{\mathrm{Planck}})^{2}$ is the gravitational constant and $\Lambda$ is the cosmological constant.

Thus we have obtained a canonical deterministic quantum jump dynamics. It must be emphasized that the construction is independent of the mechanism of quantum jumps.

\subsection{Pseudorandomness due to complexity of the universe}

In order that the canonical quantum dynamics result in an effective probabilistic structure, it is necessary (and sufficient) that the sequence
\begin{equation} 
\tilde{r}=(r_{j})_{j=1}^{n},\quad r_{j}\;\;\mathrm{given\;by}\;(4.6.3)
\end{equation}
be pseudorandom. In a physical system, the pseudorandomness of this sort has its origin in the complexity of the system c[15--17,5,18]. And the universe is the most complicated system.

\subsection{Classical and quantum determinism. \\Arrow of time}

In conclusion, we point out a difference between classical and quantum determinism. Classical determinism, i.e., determinism in classical physics may be expressed by the following implication:
\begin{equation} 
\mathrm{classical\;\;determinism}\Leftrightarrow\mathrm{predictability\;\;and\;\;
retrodictability}
\end{equation}
whereas quantum determinism, i.e., determinism in quantum physics is characterized by the implication
\begin{equation} 
\mathrm{quantum\;\;determinism}\Leftrightarrow\mathrm{predictability}
\end{equation}
but
\begin{equation}
\mathrm{quantum\;\;determinism}\nLeftrightarrow\mathrm{retrodictability}
\end{equation}
It is relations (4.8.2) and (4.8.3) that give rise to the arrow of time in the quantum universe.

\newpage
\section*{Acknowledgments}

I would like to thank Alex A. Lisyansky for support and Stefan V.
Mashkevich for helpful discussions.

\end{document}